\documentclass{elsart}
\usepackage{amsmath}
\usepackage{amssymb}
\usepackage{graphicx}
\usepackage{latexsym}
\usepackage{aas_macros}
\usepackage{natbib}
\usepackage[squaren]{SIunits}
\usepackage{array}
\usepackage{tabularx}

\newcommand{\paren}[1]{\left ( #1 \right )}

\newcommand{\parenfrac}[2]{\paren{\frac{#1}{#2}}}

\newcommand{\differd}[1]{\textrm{d}^{#1}}
\newcommand{\differ}[1]{\differd{}#1}
\newcommand{\sdiffer}[1]{\, \differ{#1}}

\newcommand{\differn}[2]{\differd{#1}#2}

\newcommand{\deriv}[2]{\frac{\differ{#1}}{\differ{#2}}}
\newcommand{\derivn}[3]{\frac{\differn{#1}{#2}}{\differ{{#3}^{#1}}}}

\addunit{\cm}{\centi\metre}

\addunit{\dyne}{dyn}
\addunit{\erg}{erg}

\addunit{\gauss}{gauss}

\addunit{\persqcm}{\per \cm \squared}
\addunit{\persqcmnp}{\cm \rpsquared}

\addunit{\percubiccm}{\per \cm \cubed}
\addunit{\percubiccmnp}{\cm \rpcubed}

\addunit{\grampersqcm}{\gram \persqcm}
\addunit{\grampersqcmnp}{\gram \usk \persqcmnp}
\addunit{\grampercubiccm}{\gram \percubiccm}
\addunit{\grampercubiccmnp}{\gram \usk \percubiccmnp}

\addunit{\ergpercubiccm}{\erg \percubiccm}
\addunit{\ergpercubiccmnp}{\erg \usk \percubiccmnp}

\addunit{\molpercubiccm}{\mole \percubiccm}
\addunit{\molpercubiccmnp}{\mole \usk \percubiccmnp}

\addunit{\cmpersec}{\cm \per \second}
\addunit{\cmpersecnp}{\cm \usk \reciprocal \second}

\addunit{\cmpersecsq}{\cm \per \second \squared}
\addunit{\cmpersecsqnp}{\cm \usk \second \rpsquared}

\addunit{\gramcmpersec}{\gram \usk \cmpersec}
\addunit{\gramcmpersecnp}{\gram \usk \cmpersecnp}

\addunit{\gramsqcmpersec}{\gram \usk \cm \squared \per \second}
\addunit{\gramsqcmpersecnp}{\gram \usk \cm \squared \second \rpsquared}

\addunit{\yyear}{yr} 
\addunit{\MBU}{MBU} 

\addunit{\jansky}{Jy}

\addunit{\magnitude}{mag}

\newcommand{\mySun}{\odot}

\addunit{\Msol}{\ensuremath{\mathrm{M}_{\mySun}}}
\addunit{\Rsol}{\ensuremath{\mathrm{R}_{\mySun}}}
\addunit{\Lsol}{\ensuremath{\mathrm{L}_{\mySun}}}
\addunit{\Zsol}{\ensuremath{\mathrm{Z}_{\mySun}}}

\addunit{\Mearth}{\ensuremath{\mathrm{M}_{\earth}}}
\addunit{\Rearth}{\ensuremath{\mathrm{R}_{\earth}}}

\addunit{\Mjup}{\ensuremath{\mathrm{M}_{J}}}
\addunit{\Rjup}{\ensuremath{\mathrm{R}_{J}}}

\addunit{\AU}{au}
\addunit{\lightyear}{ly}
\addunit{\parsec}{pc}

\addunit{\cmsqpergramnp}{\centi\metre\squared\usk\reciprocal\gram}

\newcommand{\vinf}[1]{v_{\infty}^{#1}}
\newcommand{\rhoinf}[1]{\rho_{\infty}^{#1}}
\newcommand{\cinf}[1]{c_{\infty}^{#1}}

\newcommand{\zetaHL}[1]{\zeta_{\text{HL}}^{#1}}
\newcommand{\zetaBH}[1]{\zeta_{\text{BH}}^{#1}}

\newcommand{\mdotHL}[1]{\dot{M}_{\text{HL}}^{#1}}
\newcommand{\mdotBH}[1]{\dot{M}_{\text{BH}}^{#1}}

\begin{document}

\runauthor{R.~G.~Edgar}

\begin{frontmatter}

\title{A Review of Bondi--Hoyle--Lyttleton Accretion}

\author[Stockholm]{Richard Edgar}
\address[Stockholm]{Stockholms observatorium, AlbaNova universitetscentrum, SE-106 91, Stockholm, Sweden}

\ead{rge21@astro.su.se}

\begin{abstract}
If a point mass moves through a uniform gas cloud, at what rate does it accrete
material?
This is the question studied by Bondi, Hoyle and Lyttleton.
This paper draws together the work performed in this area since the
problem was first studied.
Time has shown that, despite the simplifications made, Bondi, Hoyle and Lyttleton
made quite accurate predictions for the accretion rate.
Bondi--Hoyle--Lyttleton accretion has found application in many fields of astronomy,
and these are also discussed.
\end{abstract}


\end{frontmatter}


\section{Introduction}

In its purest form, Bondi--Hoyle--Lyttleton accretion concerns the supersonic
motion of a point mass through a gas cloud.
The cloud is assumed to be free of self-gravity, and to be uniform at
infinity.
Gravity focuses material behind the point mass, which can then
accrete some of the gas.
This problem has found applications in many areas of astronomy,
and this paper is an attempt to address the lack of a general
review of the subject.

I start with a short summary of the original work of Bondi, Hoyle
and Lyttleton, followed by a discussion of the numerical simulations
performed.
Some issues in Bondi--Hoyle--Lyttleton accretion are discussed, before
a brief summary of the fields in which the geometry has proved useful.

\section{Basics}

This section is somewhat pedagogical in nature, containing a brief summary
of the work of Bondi, Hoyle and Lyttleton.
Readers familiar with the basic nature of Bondi--Hoyle--Lyttleton accretion
may wish to skip this section.

\subsection{The Analysis of Hoyle \& Lyttleton}

\begin{figure}
\centering
\includegraphics[scale=1.0]{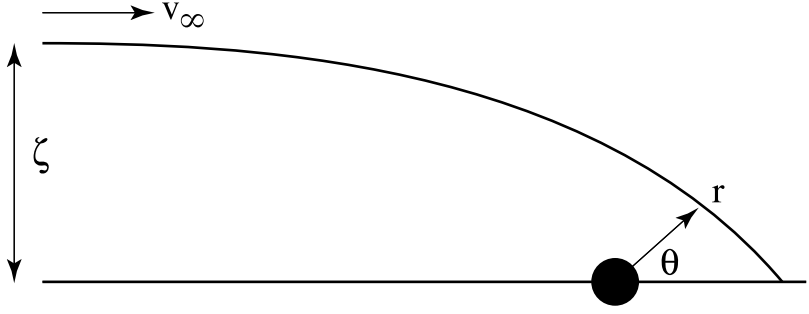}
\caption{Sketch of the Bondi--Hoyle--Lyttleton accretion geometry}
\label{fig:BHgeometry}
\end{figure}

\citet{1939PCPS.34..405} considered accretion by a star moving at
a steady speed through an infinite gas cloud.
The gravity of the star focuses the flow into a wake which it then
accretes.
The geometry is sketched in figure~\ref{fig:BHgeometry}.

\citeauthor{1939PCPS.34..405} derived the accretion rate in the following
manner:
Consider a streamline with impact parameter $\zeta$.
If this follows a ballistic orbit (it will if pressure effects are negligible),
then we can apply conventional orbit theory.
We have
\begin{eqnarray}
\ddot{r} - r \dot{\theta}^2 & = & - \frac{GM}{r^2} \\
r^2 \dot{\theta} & = & \zeta \vinf{}
\end{eqnarray}
in the radial and polar directions respectively.
Note that the second equation expresses the conservation
of angular momentum.
Setting $h = \zeta \vinf{}$ and making the usual
substitution $u = r^{-1}$, we may rewrite the first equation as
\begin{equation}
\derivn{2}{u}{\theta} + u = \frac{GM}{h^2}
\end{equation}
The general solution is $u = A \cos \theta + B \sin \theta + C$
for arbitrary constants $A$, $B$ and $C$.
Substitution of this general solution immediately shows that
$C = GM/h^2$.
The values of $A$ and $B$ are fixed by the boundary conditions
that $u \rightarrow 0$ (that is, $r \rightarrow \infty$) as
$\theta \rightarrow \pi$, and that
\begin{equation*}
\dot{r} = - h \deriv{u}{\theta} \rightarrow - \vinf{} \text{ as } \theta \rightarrow \pi
\end{equation*}
These will be satisfied by
\begin{equation}
u = \frac{GM}{h^2}\paren{1 + \cos \theta} - \frac{\vinf{}}{h} \sin \theta
\label{eq:BHOrbitEquation}
\end{equation}

Now consider when the flow encounters the $\theta = 0$ axis.
As a first approximation, the $\theta$ velocity will go to zero
at this point.
The radial velocity will be $\vinf{}$ and the radius of the streamline
will be given by
\begin{equation}
\frac{1}{r} = \frac{2 G M}{h^2}
\end{equation}
Assuming that material will be accreted if it is bound to the star
we have
\begin{equation*}
\frac{1}{2}\vinf{2} - \frac{GM}{r} < 0
\end{equation*}
or
\begin{equation}
\zeta < \zetaHL{} = \frac{2 G M}{\vinf{2}}
\label{eq:HoyleLyttletonRadiusDefine}
\end{equation}
which defines the critical impact parameter, known as the
Hoyle--Lyttleton radius.
Material with an impact parameter smaller than this value
will be accreted.
The mass flux is therefore
\begin{equation}
\mdotHL{} = \pi \zetaHL{2} \vinf{} \rhoinf{} = \frac{4 \pi G^2 M^2 \rhoinf{}}{\vinf{3}}
\label{eq:HoyleLyttletonAccRateDefine}
\end{equation}
which is known as the Hoyle--Lyttleton accretion rate.

\subsection{Analytic Solution}

The Hoyle--Lyttleton analysis contains no fluid effects, which makes it ripe
for analytic solution.
This was performed by \citet{1979SvA....23..201B}, who derived the following
solution for the flow field:
\begin{eqnarray}
v_{r} & = & - \sqrt{ \vinf{2} + \frac{2 G M}{r} - \frac{\zeta^2 \vinf{2}}{r^2}}
\label{eq:BHanalyticvr} \\
v_{\theta} & = & \frac{\zeta \vinf{}}{r}
\label{eq:BHanalyticvtheta} \\
r & = & \frac{ \zeta^2 \vinf{2}}{GM(1+\cos \theta) + \zeta \vinf{2} \sin \theta}
\label{eq:BHanalyticr} \\
\rho & = & \frac{\rhoinf{} \zeta^2}{r \sin \theta ( 2 \zeta - r \sin \theta )}
\label{eq:BHanalyticrho}
\end{eqnarray}
The first three equations are fairly straightforward, and follow (albeit tediously)
from the orbit solution given above.
The equation for the density is rather less pleasant, and involves solving
the steady state gas continuity equation under conditions of
axial symmetry.

Equation~\ref{eq:BHOrbitEquation} may be rewritten into the
form
\begin{equation}
r = \frac{r_0}{1 + e \cos( \theta - \theta_0)}
\end{equation}
where $e$ is the eccentricity of the orbit, $r_0$ is the semi-latus
rectum, and $\theta_0$ is the periastron angle.
These quantities may be expressed as
\begin{eqnarray}
\theta_0 & = & \tan^{-1} \parenfrac{\zeta \vinf{2}}{GM} \\
e        & = & \sqrt{1 + \frac{\zeta^2 \vinf{4}}{G^2 M^2}} \\
r_0      & = & \frac{\zeta^2 \vinf{2}}{GM}
\end{eqnarray}
which may be useful as an alternative form to equation~\ref{eq:BHanalyticr}.

Note that these equations do not follow material down to the accretor.
Accretion is assumed to occur through an infinitely thin, infinite density
column on the $\theta=0$ axis.
This is not physically consistent with the ballistic assumption, since
it would not be possible to radiate away the thermal energy released
as the material loses its $\theta$ velocity.
Even with a finite size for the accretion column, a significant trapping
of thermal energy would still be expected.
For now we shall neglect this effect.

\subsection{The Analysis of Bondi and Hoyle}
\label{sec:BondiHoyleAnalysis}

\begin{figure}
\centering
\includegraphics[scale=1.0]{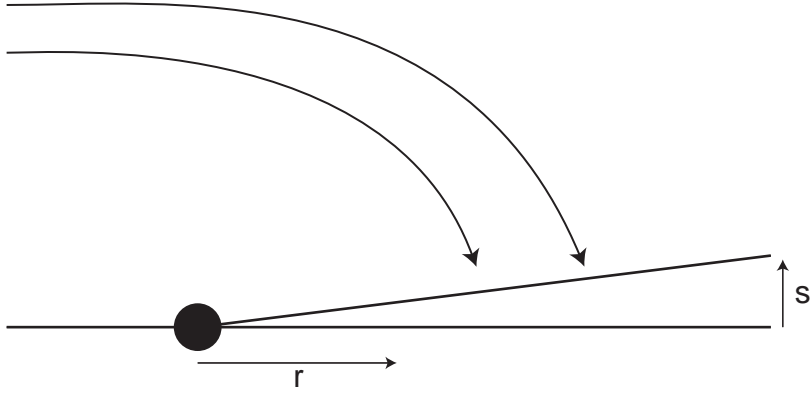}
\caption{Sketch of the geometry for the Bondi--Hoyle analysis}
\label{fig:BondiHoyleAnalysis}
\end{figure}

\citet{1944MNRAS.104..273B} extended the analysis to include
the accretion column (the wake following the point mass on the
$\theta = 0$ axis).
We will now follow their reasoning,
and show that this suggests
that the accretion rate could be as little as half the value
suggested in equation~\ref{eq:HoyleLyttletonAccRateDefine}.
Figure~\ref{fig:BondiHoyleAnalysis} sketches the quantities
we shall use.

From the orbit equations, we know that material encounters
the $\theta=0$ axis at
\begin{equation*}
r = \frac{\zeta^{2} \vinf{2}}{2 G M}
\end{equation*}
This means that the mass flux arriving in the distance
$r$ to $r + \differ{r}$ is given by
\begin{equation}
2 \pi \zeta \sdiffer{\zeta} \cdot \rhoinf{} \vinf{}
=
\frac{2 \pi G M \rhoinf{}}{\vinf{}} \differ{r}
=
\Lambda \differ{r}
\label{eq:LambdaDefine}
\end{equation}
which defines $\Lambda$.
Note that it is independent of $r$.
The transverse momentum flux in the same interval is given
by
\begin{equation*}
\Lambda \cdot v_{\theta}(\theta = 0) \cdot \frac{1}{2 \pi s}
\end{equation*}
which is the mass flux, multiplied by the transverse velocity,
divided over the approximate area of the wake.
Applying the orbit equations once more, and noting that
a momentum flux is the same as a pressure, we find
\begin{equation}
P_{s} \approx \frac{\Lambda}{2 \pi s} \sqrt{\frac{2 G M}{r}}
\label{eq:WakePressure}
\end{equation}
as an estimate of the pressure in the wake.
The longitudinal pressure force is therefore
\begin{equation*}
\differ{(\pi s^{2} P_{s})} = \Lambda \sqrt{\frac{GM}{2}} \differ{\parenfrac{s}{\sqrt{r}}}
\end{equation*}
Material will take a time of about $r/\vinf{}$ to fall onto the
accretor from the point it encounters the axis.
This means that we can use the accretion rate to estimate the mass
per unit length of the wake, $m$, as
\begin{equation}
m \approx \Lambda \frac{G M}{\vinf{3}}
\label{eq:massperunitlengthdefine}
\end{equation}
This makes the gravitational force per unit length
\begin{equation*}
F_{\text{grav}}
=
\frac{G M m \sdiffer{r}}{r^{2}}
\approx
\Lambda \frac{G^{2} M^{2}}{\vinf{3}} \cdot \frac{\differ{r}}{r^{2}}
\end{equation*}
For accreting material, we must have $r \sim G M \vinf{-2}$.
If we also assume that the wake is thin ($s \ll r$) and
roughly conical ($\differ{s} / s \approx \differ{r} / r$), then
taking the ratio of the pressure and gravitational forces, we find
that pressure force is much less than the gravitational force.
We can therefore neglect the gas pressure in the wake.

The mass per unit length of the wake, $m$, was introduced above.
If we assume the mean velocity in the wake is $v$, we can write
two conservation laws, for mass and momentum:
\begin{eqnarray}
\deriv{}{r} ( m v ) & = & \Lambda \label{eq:WakeConsMass} \\
\deriv{}{r} (m v^2) & = & \Lambda \vinf{} - \frac{G M m}{r^2} \label{eq:WakeConsMom}
\end{eqnarray}
Recall that $\Lambda \vinf{}$ is the momentum supply into the wake, since
$\dot{r} = \vinf{}$ on axis for all streamlines.
We can declutter these equations by introducing dimensionless
variables for $m$, $r$ and $v$:
\begin{eqnarray}
v & = & \vinf{} \nu \\
r & = & \frac{G M}{\vinf{2}} \chi \\
m & = & \frac{\Lambda G M}{\vinf{3}} \mu
\end{eqnarray}
Note that $\chi=2$ corresponds to material arriving from the streamline
characterised by $\zetaHL{}$.
Substituting these definitions into equations~\ref{eq:WakeConsMass}
and~\ref{eq:WakeConsMom}, we obtain
\begin{eqnarray}
\deriv{}{\chi} ( \mu \nu ) & = & 1 \label{eq:ScaledWakeMass} \\
\deriv{}{\chi} ( \mu \nu^{2} ) & = & 1 - \mu \chi^{-2} \label{eq:ScaledWakeMom}
\end{eqnarray}
We shall now analyse the behaviour of these equations.

We can integrate equation~\ref{eq:ScaledWakeMass} to yield
\begin{equation}
\mu \nu = \chi - \alpha
\label{eq:munuintegrate}
\end{equation}
for some constant $\alpha$.
Since $\mu$ is a scaled mass (and hence always positive), we see that
the scaled velocity ($\nu$) changes sign when $\chi = \alpha$.
That is, $\alpha$ is the stagnation point.
Material for $\chi < \alpha$ will accrete, so knowing $\alpha$ will
tell us the accretion rate (since the accretion rate will be $\Lambda r_0$
where $r_0$ is the value of $r$ corresponding to $\alpha$).
By writing $\mu \nu^{2} = \mu \nu \cdot \nu$, we can use
equation~\ref{eq:munuintegrate} to rewrite equation~\ref{eq:ScaledWakeMom}
as
\begin{equation}
\nu \deriv{\nu}{\chi} = \frac{\nu(1-\nu)}{\chi - \alpha} - \frac{1}{\chi^{2}}
\label{eq:ScaledWakeMom2}
\end{equation}
This has not obviously improved matters, but we can now study the
general behaviour of the function, without trying to solve it.
First we need some boundary conditions.
These are as follows:
\begin{itemize}
\item $\nu \rightarrow 1$ as $\chi \rightarrow \infty$ \\
      Which means that $v \rightarrow \vinf{}$ at large radii

\item $\nu = 0$ at $\chi = \alpha$ \\
      The stagnation point

\item \begin{equation*}\deriv{\nu}{\chi} > 0 \text{ Everywhere}\end{equation*} \\
      The velocity must be a monotonic function.
      This is physically reasonable, if we are to avoid unusual flow patterns
\end{itemize}
The first two conditions can be satisfied for any value of $\alpha$.
Fortunately, the third implies as restriction.
The next set of manipulations may seem a little obscure at first, but they do lead
in the desired direction.

Substitute $\xi = \alpha^{-1} \chi$.
Equation~\ref{eq:ScaledWakeMom2} then reads
\begin{equation}
\nu \deriv{\nu}{\xi} = \frac{\nu (\nu-1)}{\xi -1} - \frac{1}{\alpha \xi^{2}}
\label{eq:ScaledWakeMom3}
\end{equation}
Now, suppose the derivative is zero.
This leads to the condition
\begin{equation*}
\nu^{2} - \nu + \frac{1}{\alpha \xi^{2}}(x-1) = 0
\end{equation*}
or, one application of the quadratic roots formula later:
\begin{equation}
\nu = \frac{1}{2} \pm \sqrt{\frac{1}{4} -  \frac{1}{\alpha \xi^{2}}(x-1)}
\label{eq:IsoclineDefine}
\end{equation}
Since $\nu$ ultimately represents a physical quantity (the velocity), it's
obviously desirable that it remain real.
We therefore need to look at when the discriminant can become zero.
This is another quadratic equation, leading to
\begin{equation*}
\xi = \frac{2}{\alpha} \paren{1 \pm \sqrt{1 - \alpha}}
\end{equation*}
which means that \emph{something} must happen when $\alpha=1$.
To determine this `something,' it is best to plot
equation~\ref{eq:IsoclineDefine}.

\begin{figure}
\centering
\includegraphics[scale=0.6]{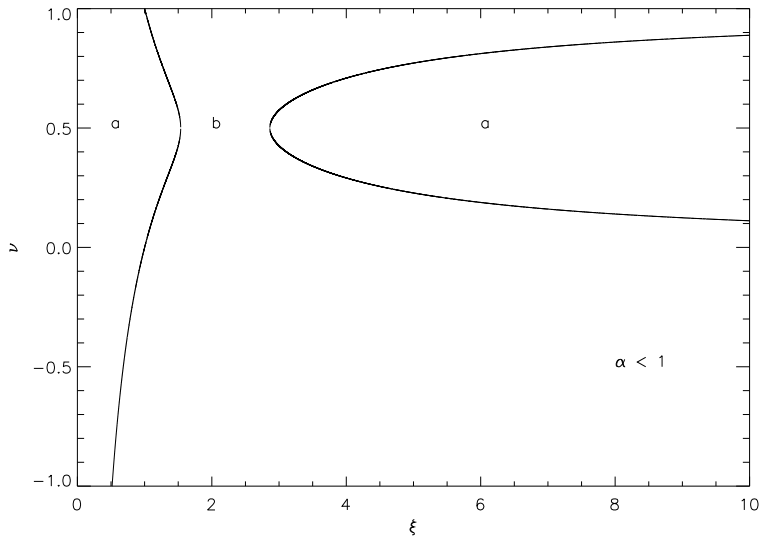}
\caption{Curves where $\deriv{\nu}{\xi}=0$ for $\alpha<1$.
In the regions marked `a,' the derivative is greater than zero.
It is less than zero in the `b' region}
\label{fig:BHisoclineAlphalt1}
\end{figure}

\begin{figure}
\centering
\includegraphics[scale=0.6]{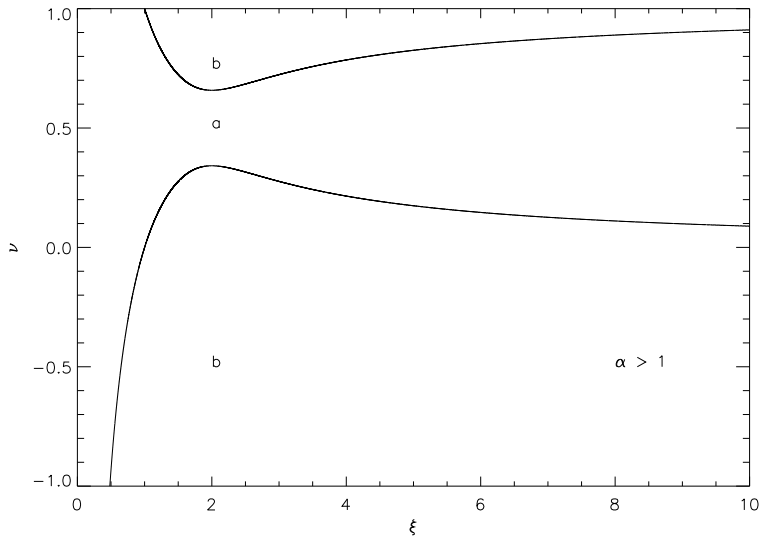}
\caption{Curves where $\deriv{\nu}{\xi}=0$ for $\alpha>1$.
In the region marked `a,' the derivative is greater than zero.
It is less than zero in the `b' regions}
\label{fig:BHisoclineAlphagt1}
\end{figure}

Figures~\ref{fig:BHisoclineAlphalt1} and~\ref{fig:BHisoclineAlphagt1}
demarcate the regions where $\deriv{\nu}{\xi}$ changes sign, as
dictated by equation~\ref{eq:IsoclineDefine}.
These are \emph{not} possible solutions for $\nu$.
However, any suitable solution for $\nu$ must remain within
the region marked `a' if it is to remain monotonic and increasing.
This is only possible when $\alpha > 1$.
Unwinding our rescaled variables, we see that an $\alpha$ value
of unity puts the stagnation point halfway between the accretor
and the original value of \citeauthor{1939PCPS.34..405}.
This in turn implies a minimum accretion rate of $0.5 \mdotHL{}$.

Again, I would like to remind the reader that the flow has been
assumed to remain isothermal with negligible gas pressure
throughout this discussion.
This assumption is likely to be violated in the wake, where
densities will be high and radiative heat loss inefficient.
At the very least, thermal effects should be important close
to the stagnation point in the wake.
\citet{2000ApJ...541..821H} details an analysis similar to that
given above, but with a pressure term included.
The value of $\alpha$ (which \citeauthor{2000ApJ...541..821H} calls
$x_0$) was found to lie between 0.6 and 3.5 for flows which
were supersonic at infinity and subject to Newtonian physics
(the polytropic and adiabatic indices were also free parameters
in this analysis).

Will the flow be stable?
\citeauthor{1944MNRAS.104..273B} asserted that
if $\alpha > 2$ (note that $\alpha=2$ gives the solution of
\citeauthor{1939PCPS.34..405}), then the wake would become
unstable to perturbations which preserve axial symmetry.
However, later analysis by \citet{1977MNRAS.180..491C} suggested
that the wake should be unstable, regardless of the value
of $\alpha$.
Subsequent numerical simulations and analytic work have shown
that Bondi--Hoyle--Lyttleton flow is far from stable, and
we will discuss the subject in section~\ref{sec:IssuesStability}.

\subsection{Connection to Bondi Accretion}

\citet{1952MNRAS.112..195B} studied spherically symmetric
accretion onto a point mass.
The analysis shows (see e.g. \citet{2002apa..book.....F})
that a Bondi radius may be defined as
\begin{equation}
r_{\text{B}} = \frac{GM}{c_{\text{s}}^{2}(r_{\text{B}})}
\label{eq:BondiRadiusDefine}
\end{equation}
Flow outside this radius is subsonic, and the density
is almost uniform.
Within it, the gas becomes supersonic and moves towards a freefall solution.
The similarities between equations~\ref{eq:HoyleLyttletonRadiusDefine}
and~\ref{eq:BondiRadiusDefine}
led \citeauthor{1952MNRAS.112..195B} to propose an interpolation
formula:
\begin{equation}
\dot{M} = \frac{2 \pi G^2 M^2 \rhoinf{}}{(\cinf{2} + \vinf{2})^{3/2}}
\label{eq:BondiHoyleAccRateInterpolate}
\end{equation}
This is often known as the Bondi--Hoyle accretion rate.
On the basis of their numerical calculations, \citet{1985MNRAS.217..367S}
suggest that equation~\ref{eq:BondiHoyleAccRateInterpolate} should
acquire an extra factor of two, to become
\begin{equation}
\mdotBH{} = \frac{4 \pi G^2 M^2 \rhoinf{}}{(\cinf{2} + \vinf{2})^{3/2}}
\label{eq:BondiHoyleAccRateDefine}
\end{equation}
which then matches the original Hoyle--Lyttleton rate as the sound
speed becomes insignificant.
The corresponding $\zetaBH{}$ is formed by analogy with
equation~\ref{eq:HoyleLyttletonAccRateDefine}.

Nomenclature in this field can be a little confused.
When papers refer to `Bondi--Hoyle accretion rates,' they may mean
equation~\ref{eq:HoyleLyttletonAccRateDefine}, \ref{eq:BondiHoyleAccRateInterpolate}
or~\ref{eq:BondiHoyleAccRateDefine}.
In this review, I shall refer to pressure-free flow as `Hoyle--Lyttleton'
accretion and use $\mdotHL{}$ and $\zetaHL{}$.
When there is gas pressure, I will talk about `Bondi--Hoyle accretion' and
use $\mdotBH{}$ and $\zetaBH{}$, in the sense defined by
equation~\ref{eq:BondiHoyleAccRateDefine}.
I shall use `Bondi--Hoyle--Lyttleton' accretion to refer to the
problem in general terms.

\section{Numerical Simulations}

In the previous section, I outlined the basic theory
behind Bondi--Hoyle--Lyttleton accretion.
This lead to elegant predictions for the accretion rate,
as given by equations~\ref{eq:HoyleLyttletonAccRateDefine}
and~\ref{eq:BondiHoyleAccRateDefine}.
However, reaching these equations required a lot of simplifying
assumptions, so necessitating further investigation.
The intractability of the equations of fluid dynamics requires
a numerical approach to the problem.

In a break with tradition, I shall start this section with the
answer, and then give more detailed citations to examples.

\subsection{Summary}

Do the equations of~\citeauthor{1979SvA....23..201B} provide
a good description of Bondi--Hoyle--Lyttleton flow?
The answer is `No.'

\begin{figure}
\center
\includegraphics[scale=0.8]{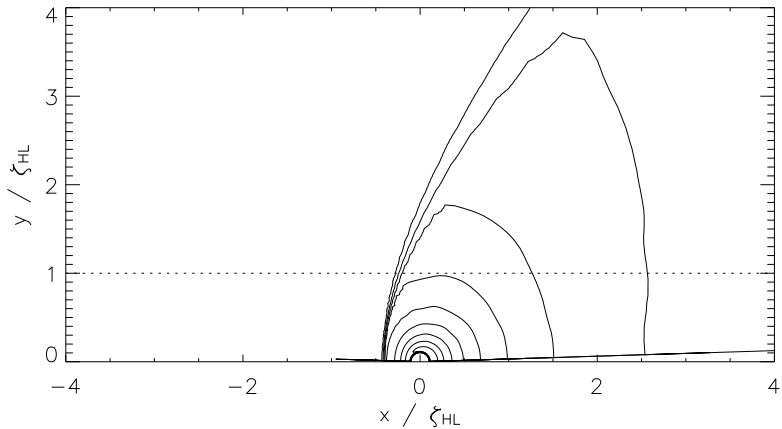}
\caption{Density contours for a sample Bondi--Hoyle--Lyttleton simulation.
The flow had $\mathcal{M}=1.4$, an accretor radius of $0.1 \zetaHL{}$ and
the equation of state was adiabatic with $\gamma=5/3$.
The contours are logarithmically spaced over a decade of density.
The dotted line indicates $\zetaHL{}$.
The flow is incident from the left}
\label{fig:PlainBHflowDensityRun2}
\end{figure}

\begin{figure}
\center
\includegraphics[scale=0.8]{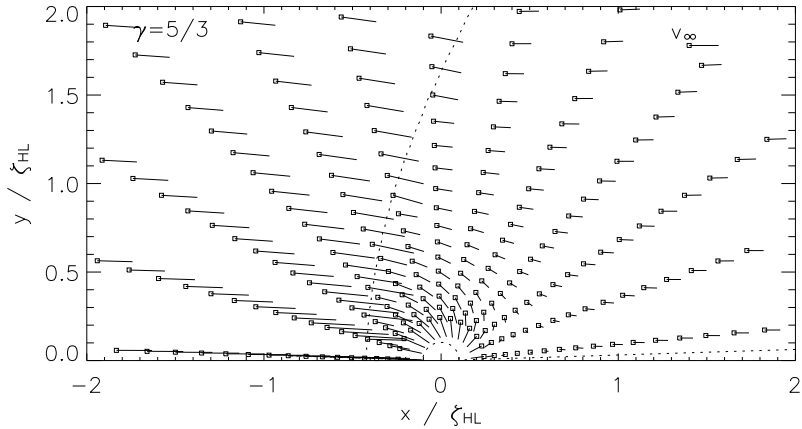}
\caption{Velocity field corresponding to the densities shown in figure~\ref{fig:PlainBHflowDensityRun2}.
The approximate position of the bow shock is marked with a dotted line}
\label{fig:PlainBHflowVelocityRun2}
\end{figure}

In the absence of other effects, three numbers parameterise
Bondi--Hoyle--Lyttleton flow:
\begin{itemize}
\item The Mach number, $\mathcal{M}$
\item The size of the accretor, in units of $\zetaHL{}$
\item The $\gamma$ value of the gas
\end{itemize}
Figure~\ref{fig:PlainBHflowDensityRun2} shows sample density
contours for a flow with $\mathcal{M}=1.4$, an accretor radius of
$0.1 \zetaHL{}$ and $\gamma=5/3$.
This particular simulation was axisymmetric.
A bow shock has formed on the upstream side.
The corresponding velocity field is plotted in
figure~\ref{fig:PlainBHflowVelocityRun2}.
Downstream of the shock, material flows almost radially onto the
accretor, in marked contrast to the analytic solution of
equations~\ref{eq:BHanalyticvr} to~\ref{eq:BHanalyticrho}.

\begin{figure}
\centering
\includegraphics[scale=0.6]{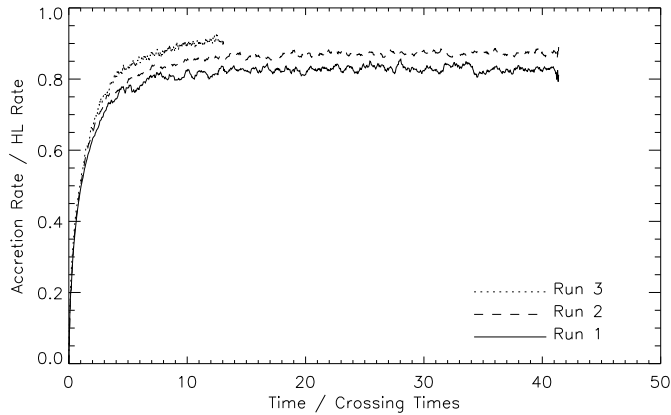}
\caption{Accretion rates for plain Bondi--Hoyle--Lyttleton flow.
The crossing time corresponds to $\zetaHL{}$}
\label{fig:PlainBHflowMdotRuns123}
\end{figure}

But what of the accretion rate?
Figure~\ref{fig:PlainBHflowMdotRuns123} shows the accretion
rates obtained for three simulations.
Although the dimensionless parameters were kept the same, the
physical scales and grid resolution varied.
Figures~\ref{fig:PlainBHflowDensityRun2} and~\ref{fig:PlainBHflowVelocityRun2}
were taken from run~2.
The accretion rates achieved are quite close to the value of $\mdotHL{}$
predicted for the flow (this value is substantially larger than the corresponding
$\mdotBH{}$).
Despite the simplifications made, the work of Bondi, Hoyle and Lyttleton has been
largely vindicated.
In the remainder of the section, I shall cite places in the literature where
further simulations of Bondi--Hoyle--Lyttleton flow may be found.

\subsection{Examples in the Literature}

\citeauthor{1971MNRAS.154..141H} computed numerical solutions of
Bondi--Hoyle--Lyttleton flow in two papers written in \citeyear{1971MNRAS.154..141H}
and \citeyear{1979MNRAS.188...83H}.
The accretion rate suggested by equation~\ref{eq:BondiHoyleAccRateDefine}
agreed well with that observed, despite the flow pattern being
rather different.
\citeauthor{1971MNRAS.154..141H} studied flows which were not
very supersonic and were non-isothermal.
A bow shock formed upstream of the accretor.
Upstream of the shock, the flow pattern was very close to the original
ballistic approximation.
Downstream, the gas flowed almost radially towards the point mass.
A summary of early calculations of Bondi--Hoyle--Lyttleton flow may be found in
\citet{1985MNRAS.217..367S}.
The calculations in this paper are in broad agreement with earlier
work, but some differences are noted and attributed to resolution
differences.

More recently, a series of calculations in three dimensions have
been performed by \citeauthor{1994ApJ...427..342R} in a series
of papers
\citep{1994ApJ...427..342R,1994ApJ...427..351R,1994A&AS..106..505R,1995A&AS..113..133R,1996A&A...311..817R}.
This series of papers used a code based on nested grids, to permit
high resolution at minimal computational cost.
\citet{1994ApJ...427..342R} details the code, and presents simulations
of Bondi accretion (where the accretor is stationary).
Bondi--Hoyle--Lyttleton flow was considered in \citet{1994ApJ...427..351R}.
The flow of gas with $\mathcal{M}=3$ and $\gamma = 5/3$ past an accretor
of varying sizes ($0.01 < r/\zetaBH{} < 10$) was studied.
For accretors substantially smaller than $\zetaBH{}$, the accretion rates
obtained were in broad agreement with theoretical predictions.
The flow was found to have quiescent and active phases, with smaller
accretors giving larger fluctuations.
However, these fluctuations were far less violent than the `flip-flop'
instability observed in 2D simulations (see below).
\citet{1994A&AS..106..505R} extended these simulations to cover a range
of Mach numbers, finding that higher Mach numbers tended to give lower
accretion rates (down to the original interpolation formula of
equation~\ref{eq:BondiHoyleAccRateInterpolate}).
\citeauthor{1995A&AS..113..133R} studied the flow of a gas with
$\gamma = 4/3$ in the \citeyear{1995A&AS..113..133R} paper, finding accretion
rates comparable with the theoretical results.
Small accretors and fast flows were required before any instabilities
appeared in the flow.
Nearly isothermal flow was considered in \citet{1996A&A...311..817R}.
The accretion rates were slightly higher than the theoretical values (except
for the smaller accretors), and the shock moved back to become a tail
shock.
The oscillations in the flow were less violent still.

The reason for the formation of the bow shock is straightforward - the
rising pressure in the flow.
As shown by equation~\ref{eq:BHanalyticrho}, the flow is compressed
as it approaches the accretor.
This compression will increase the internal pressure of the flow,
eventually causing a significant disruption.
At this point, the shock will form.
This interpretation is consistent with the behaviour observed in
simulations, where decreasing $\gamma$ moves the shock back towards
the accretor.
However, the precise location of the shock does not seem to be a
strong function of the Mach number
(cf the papers of \citeauthor{1996A&A...311..817R}).

\section{Issues in Bondi--Hoyle--Lyttleton Flow}
\label{sec:Issues}

In this section I shall discuss some issues relating to
Bondi--Hoyle--Lyttleton flow which are of particular interest.

\subsection{The Drag Force}

The simple idealisation of Bondi--Hoyle--Lyttleton flow cannot persist
for long.
The accretor is not only increasing its mass - it is accumulating
momentum as well.
Eventually, it should be accelerated to being co-moving with
the gas flow.
A full calculation is not straightforward, but dimensional considerations
suggest
\begin{equation}
\dot{M} \vinf{} \sim M \dot{v}_{\infty} = F_{\text{drag}}
\label{eq:RoughDrag}
\end{equation}
Please note, that the $\sim$ in this equation is \emph{very}
approximate.
However, equation~\ref{eq:RoughDrag} suggests that the
accreting body will be brought to rest with respect to
the flow on the mass doubling timescale.
This is obviously a problem if the accretor is to change its
mass appreciably.
As we shall see later, this has lead to most research into
Bondi--Hoyle--Lyttleton accretion being concentrated into the
study of binaries.
In such cases, the momentum difference can be `paid' by a change
in orbit.

The drag does not originate as a form of `wind' resistance pressing
directly on the accretor.
This is for two reasons
\begin{itemize}
\item The momentum deposited by the accretion column will be far larger
\item Mathematically, the accretor is a point anyway
\end{itemize}
Instead, the drag arises from the gravitational focusing of
material behind the accretor.
Since more material is present on the downstream side, the gravitational
attraction of the downstream side is larger,\footnote{This is ignoring
the mathematical impossibility of an infinite, uniform medium for the
unperturbed case} and exerts a drag force.
\citet{1943ApJ....97..255C} was the first to consider this problem - called
`dynamical friction' - for a collisionless fluid.
An extended (and more recent) discussion of the problem is given by
\citet{1987gady.book.....B}.
\citet{1964SvA.....8...23D} discussed the problem for a gaseous medium.
The matter of drag is also mentioned by \citet{1971ApJ...165....1R}, who
propose
\begin{equation}
F_{\text{drag}} = \dot{M} \vinf{} \ln \parenfrac{b_{\text{max}}}{b_{\text{min}}}
\end{equation}
where $b_{\text{min}}$ and $b_{\text{max}}$ are cut-off radii for the
gravitational force.
\citet{1978MNRAS.182..371Y} suggested that a suitable value for the outer cut-off
for a flow with pressure would be the point where the pressure in the wake
became equal to the background pressure (the inner cut-off radius is usually taken
to be the radius of the accretor itself).

Values for the drag force given by \citet{1985MNRAS.217..367S,1993A&A...274..955S,1994ApJ...427..351R}
suggest that the drag force is no more than a factor of ten larger
than the crude estimate of equation~\ref{eq:RoughDrag}.
The precise drag value has a tendency to fluctuate anyway - Bondi--Hoyle--Lyttleton
flow is not stable.

\subsection{Flow Stability}
\label{sec:IssuesStability}

Even in the axisymmetric case, there is no particular reason to believe
that Bondi--Hoyle--Lyttleton flow should be stable.
The binding energy test of equation~\ref{eq:HoyleLyttletonRadiusDefine}
is made for gas flowing \emph{away} from the accretor.
If this material is going to be accreted, it needs to turn around
somehow and fall towards the point mass.
This must happen in some sort of accretion column, of the type
first considered by \citet{1944MNRAS.104..273B}.
As noted above, the work of \citet{1977MNRAS.180..491C} found that
this wake should be unstable.
A `shock cone' must surround the wake, in which the flow loses
its $\theta$ velocity before it encounters the axis (see also
\citet{1977ApJ...213..200W,1977ApJ...213..208W}).
The high densities expected for the wake mean that this shock is likely
to heat the gas.
Gas pressure could then be expected to drive oscillations close to the
stagnation point.
Bondi--Hoyle--Lyttleton flow around small accretors has been studied by
\citet{1991MNRAS.252..473K}.
This paper notes that that the `accreting body is so small that a
part of the accreting gas sometimes misses the target object and
flows towards the upstream as a jet.'
This is obviously a rather unstable condition, and leads to the
accretion flow `sloshing' back and forth around the point mass.
The accretion rate fluctuates too, although the time averaged
rate is still close to the Bondi--Hoyle value.

When the condition of axisymmetry is relaxed, even more
instabilities become possible.
\citet{1987MNRAS.226..785M,1988ApJ...335..862F,1988ApJ...327L..73T}
performed two dimensional simulations of the Bondi--Hoyle--Lyttleton
geometry, with the condition of axisymmetry relaxed, and
a density and/or velocity gradient imposed on the upstream
flow.
All three papers found that a `flip-flop' instability
resulted, with the wake oscillating back and forth in
a manner reminiscent of a von K\'{a}rm\'{a}n vortex street.
\citet{1991A&A...248..301M} suggested that the instability was
intrinsic to the accretion flow, since it was found to develop
even under the conditions first considered by
\citet{1939PCPS.34..405}.
Another detailed study of the `flip-flop' instability for the
2D case for isothermal gas is that of
\citet{1998A&A...337..311S}.
The code used was specifically designed to conserve angular
momentum and to permit very high resolution in the inner portions
of the grid.
They suggest that some of the resolution dependence of the
instability found by earlier work was due to the use of codes
which conserved \emph{linear} momentum, and caution against
the results obtained by such codes.

Such a spectacular instability naturally prompted an intense
theoretical investigation.
\citet{1990ApJ...358..545S} extended the earlier work of
\citet{1977MNRAS.180..491C} to include tangential oscillations.
The analysis is based on the assumption of a pressure free
flow, and the expressions derived for the tangential behaviour
also require the flow to be 2D and planar.
The radial instability noted by \citeauthor{1977MNRAS.180..491C} was
found to be independent of the incoming material.
Any radial oscillation in the wake would grow, although the growth
timescale was much longer than the oscillation timescales.
The tangential modes (corresponding to the `flip-flop' instability)
behaved in a similar manner.
\citeauthor{1990ApJ...358..545S} also predicted that the instability
should be milder in the 3D case.
In \citet{1991ApJ...376..750S}, a numerical study of the coupling between
the radial and tangential oscillations was made.
The radial modes, corresponding to large density and velocity fluctuations
were excited far beyond the linear regime, while the tangential
oscillations remained linear.
While the mass accretion rate showed corresponding fluctuations,
the time averaged accretion rate was similar (although smaller) than
the prediction of equation~\ref{eq:HoyleLyttletonAccRateDefine}.
\citet{1991MNRAS.253..633L} added a simple analysis of the shock cone
surrounding the wake seen in numerical simulations.
Instabilities were found in both the planar 2D and full 3D cases,
although the authors note that the instability should be milder
in the 3D case (a point also made by \citet{1990ApJ...358..545S}).

The major weakness of all simulations of the `flip-flop'
instability mentioned so far is that they fundamentally change
the geometry of the problem.
In order to simulate non-axisymmetric flow in two dimensions,
the flow has to be assumed to be planar.
This changes the shape of the accretor from a sphere to a
cylinder.
The equations of fluid dynamics are non-linear, and are notorious
for their resolute refusal to yield to a proof of solution
uniqueness \citep{NavierStokesUnique}.
There is therefore no particular reason to expect the
2D planar simulations to be characteristic of the true solution
in 3D.
The simulations of \citeauthor{1994ApJ...427..342R} suggest that
the `flip-flop' instability is an artifact of 2D planar flow.

\citet{1997A&A...320..342F,1999A&A...347..901F}
attempted to model the instabilities observed in the earlier
numerical work of \citeauthor{1994ApJ...427..342R}.
The first of these papers constructs stationary models, while the
second contains a stability analysis.
The origin of the instability was the bow shock generally seen
in numerical simulations.
This produces entropy gradients in the flow, which allows
Rayleigh--Taylor and Kelvin--Helmholtz instabilities to grow.
\citeauthor{1999A&A...347..901F} concluded that the instability
should be stronger if
\begin{itemize}
\item The shock is detached from the accretor (as is the case for higher $\gamma$ values)
\item The flow has a higher Mach number
\item The accretor is smaller
\end{itemize}
They found that the instability should be non-axisymmetric, and
start at around $\theta = \pi/2$ and close to the accretor.
\citet{2000A&A...363..174F} describe the instability as `entropic--acoustic,'
where entropy perturbations introduced by a shock propagate back
to the shock via sound waves.
These then trigger new entropy perturbations.

\subsection{Non-uniform Boundary}

Suppose the conditions at infinity are not uniform, but instead a
density and/or velocity gradient is present.
This means that the flow within a cylinder of radius $\zetaHL{}$
possesses angular momentum about the accretor.
How much of this reaches the accretor?

Early calculations
\citep{1952MNRAS.112..205D,1975A&A....39..185I,1976ApJ...204..555S,1981A&A...102...36W}
suggested that most of this angular momentum would accrete.
However, \citet{1980MNRAS.191..599D} pointed out that only
material which had lost most of its angular momentum would be able
to settle onto a small accretor.
They developed a simple analytic model for small density and velocity
gradients (if the gradients become large, then the flow ceases to
behave in the manner described by Bondi, Hoyle and Lyttleton).
\citeauthor{1980MNRAS.191..599D} found that the mass accretion rate should
be unaffected, and there should be no accretion of angular momentum.

Such confusion calls for numerical work.
\citet{1995A&A...295..108R} presented a sample 3D simulation of accretion
under such conditions.
They found that about 70\% of the angular momentum available (as calculated
by \citet{1976ApJ...204..555S}) would be accreted.
\citeauthor{1997A&A...317..793R} extended this calculation with papers in
\citeyear{1997A&A...317..793R} (covering velocity gradients)
and \citeyear{1999A&A...346..861R} (studying density gradients).
The mass accretion was not affected much, while the angular momentum
accretion rate varied between 0\% and 70\% of the value suggested
by \citet{1976ApJ...204..555S}.
Smaller accretors gave less stable flow, but none were as violently
unstable as the `flip-flop' instability observed in 2D planar simulations.
\citeauthor{1997A&A...317..793R} noted that a very small accretor will
not be able to accrete all of the angular momentum within $\zetaHL{}$, but
was unable to test such a case, due to the vast computational load involved
in such a simulation.

\subsection{Radiation Pressure}

Where there is accretion, there will be an accretion luminosity.\footnote{Barring
certain cases of finely tuned accretion onto a black hole}
Radiative feedback has the potential to alter the
Bondi--Hoyle--Lyttleton flow, and a number of workers
have studied this.

Most work has concentrated on the problem of radiative feedback
in X-ray binaries.
This is a fairly straightforward application, since the constancy
of the Thompson cross section makes the transfer problem intrinsically
grey.

\citet{1990ApJ...356..591B} simulated a compact object accreting
an O star wind (forming an X-ray binary).
Radiative heating was included (it was relevant to the radiation-driven
wind), but the radiation force of the X-rays was neglected.
The gas was assumed to be optically thin and in ionisation
equilibrium.
The wake was found to be unstable, oscillating back and forth.
Despite these, the accretion rates were broadly consistent with
that expected from a Bondi--Hoyle type analysis (some modifications
were necessary to allow for the geometry of the binary).
A later study of the same problem was made by
\citet{1991ApJ...371..696T}.
This work included the effect of radiation pressure, but the flow
was still assumed to be optically thin to electron scattering.
Radiation pressure was then negligible except in the wake.
This paper contained short section considering the accretion of
an optically thin gas subject to radiation pressure.
This led to the prediction that the flow would be unstable to
oscillations if the accretion rate exceeded one third of the Eddington
Limit.

\citet{1995A&A...297..739K} were interested in the application
of Bondi--Hoyle--Lyttleton accretion following a nova explosion in a binary.
The radiation field was simulated using flux limited diffusion, and
analytic approximations to opacity values.
Their simulations had quite a complicated model for the accretor,
including an envelope.
Radiation pressure was found to be critical to simulating the flow
(heating was included in all calculations, but the radiation pressure
was omitted from some).
For hot, optically thick flow, including radiation pressure made the
flow subsonic and substantially reduced the drag.
However, the accretion rate was found to be low (much less than
the predictions of Bondi--Hoyle theory) in all cases.

\subsection{Relativity}

If the accreting object is a neutron star or a black hole,
relativistic effects are likely to become important.
However, relativistic hydrodynamics is generally recognised to be
a non-trivial problem, and relatively little work has been done
on Bondi--Hoyle--Lyttleton accretion for relativistic flows.

An early study by \citet{1989ApJ...336..313P} found broad
agreement between the Newtonian and relativistic cases.
\citet{1998ApJ...494..297F} performed axisymmetric calculations
in a Schwarzschild metric.
The passage of time gave more powerful computers, enabling
simulations to be run with higher resolution than
\citeauthor{1989ApJ...336..313P}.
Most of the accretion rates found
were similar to the Newtonian estimates, but some were an order of
magnitude or so higher.
No signs of instability were found.
\citet{1998MNRAS.298..835F} relaxed the assumption of axisymmetry,
but still found the flow to be steady.
However, the authors note that they were unable to push the
simulation parameters very far.

\citet{1999MNRAS.305..920F} simulated Bondi--Hoyle--Lyttleton
flow onto a rotating (Kerr) black hole.
Two forms of the metric were used, to differentiate between
numerical and physical effects.
They found that rotational effects were confined to a region
close to the hole.
Their flows remained steady.

\section{Applications}

The Bondi--Hoyle--Lyttleton scenario has been applied to a
variety of problems in the years since its introduction.
I will now discuss some examples of these.
Each of these problems could be the subject of a review article
by themselves (and often have been), so I am not able to discuss
the issues involved in great depth.
However, I hope that this will provide a reasonable sample of the
wide range of areas where the Bondi--Hoyle--Lyttleton model has
proven useful.

\subsection{Binary Systems}

The problem of accretion in a binary system seems
to be the most popular application of the Bondi--Hoyle--Lyttleton
analysis.
As noted above, the inevitable drag force simply causes
the orbit to change.
The gas supply can either be from a stellar wind, or from
common envelope (CE) evolution.

Wind accretion seems to be one of the most popular applications
of the Bondi--Hoyle--Lyttleton geometry.
However, there are a number of potential complications.
For the Bondi--Hoyle--Lyttleton solution to be valid, accretion
must be driven by a wind, rather
than by Roche lobe overflow.
The work of \citet{1978ApJ...224..625P} warns that the presence
of Roche lobe overflow will substantially complicate matters,
and that allowance must be made for the possibility when
comparing theory with observations.
The orbital motion of the binary can also cause problems - 
see \citet{1993MNRAS.265..946T,1996MNRAS.280.1264T} for a
discussion.
These papers studied accretion rate in a binary, where the
wind speed was comparable to the orbital velocity.
Consequently, the flow pattern was substantially different
from that envisaged by Bondi, Hoyle and Lyttleton.
\citet{1996MNRAS.280.1264T}
found that the accretion rate for a binary is
decreased by a factor of about ten compared to the prediction
of equation~\ref{eq:BondiHoyleAccRateDefine}.
They attribute this difference to the disrupting effect of the
orbital motion.
Moving to higher mass ratios,
\citet{2004MNRAS.347..173S} modelled wind accretion onto substellar
companions of Mira variables.
They found that the mean accretion rates were generally similar to
those predicted by Bondi, Hoyle and Lyttleton.
However, the flow was highly variable - both due to instabilities
in the accretion flow and the intrinsic variability of the star.
The ultimate `wind' is that produced by an explosion.
In this case, the very high velocities involved tend to make $\zetaBH{}$
comparable to the size of the accreting body.
The work of \citet{1995A&A...297..739K} has already been mentioned.
However, the problem had been studied before - \citet{1980MNRAS.191..933M}
estimated accretion rates using modified a Bondi--Hoyle--Lyttleton formula.

\citet{1975MNRAS.172..483J} used the predictions of Bondi--Hoyle--Lyttleton
theory to derive the system parameters of Cen X-3.
Similar work was performed by \citet{1975MNRAS.172P..35E},
\citet{1975MNRAS.172..473P} and \citet{1976A&A....49..327L}.
By considering the accretion of a high velocity wind by
neutron stars, \citet{2002ApJ...571L..37P} concluded that
most of the low luminosity, hard X-ray sources known in our Galaxy
could be powered by such systems.
Modified Bondi--Hoyle--Lyttleton accretion has also been used
to study cases where a giant star's wind is being accreted by a
main sequence star.
Some examples are the work of \citet{1981ApJ...248.1043C} and
\citet{1986A&A...156..172C}.
The Bondi--Hoyle--Lyttleton geometry is a useful first approximation
to wind accretion in binary systems.
However, unless the wind speed is much greater than the orbital
velocity, the accretion rates can deviate significantly from the
simple predictions.

At the extreme end of the mass ratio scale, Bondi--Hoyle--Lyttleton
accretion has even been used to estimate accretion rates onto
a planet embedded in a disc \citep{2003ApJ...589..578N}.
Although the situation simulated was not entirely appropriate to
the original analysis, \citeauthor{2003ApJ...589..578N} point out
that it represents a \emph{maximum} possible accretion rate.
This rate turns out to be extremely high, showing the need for
higher resolution simulations of planetary accretion flows.

Common Envelope evolution occurs when two stellar cores become
embedded in a large gas envelope.
Such an envelope is typically produced when one of the members
of the binary system swells as it leaves the main sequence.
For a more detailed discussion of Common Envelope evolution itself
see, e.g. \citet{1993PASP..105.1373I}.
In such cases, accretion rates are critical for determining the
detailed evolution of the system.
In computing accretion rates, modified Bondi--Hoyle--Lyttleton
formul\ae{} are often used e.g. \citet{1989ApJ...337..849T} - see
also the review by \citet{2000ARA&A..38..113T}.

\subsection{Protostellar Clusters}

Bondi--Hoyle--Lyttleton flow is also likely to be applicable to regions of
star formation.
Although single stars will be stopped by the drag force, real stars
are generally thought to form in clusters.
Protostars and gas are trapped inside a gravitational well, and
orbit within it.
The drag will then simply cause a change in orbit - as is the case
for X-ray binaries.
Indeed, the approximation is likely to be better for protostars in
a protocluster.
This is because the orbital motion of the protostars is the `source'
of the wind, rather than a wind from a companion.
As a result, the geometry is simpler (since the orbital motion does
not have to be added to the wind velocity).
Furthermore, non-inertial forces (coriolis and centrifugal) are likely
to be far less significant.

\citet{2001MNRAS.323..785B} performed a thorough study of
accretion in a protocluster.
They simulated the evolution of a gas cloud containing many small point
masses (representing protostars).
The point masses grew by accreting the gas.
When the gas dominated the mass of the cluster, \citeauthor{2001MNRAS.323..785B}
found that the accretion was best described by tidal lobe overflow
(examining whether material was bound to the cluster or the star).
However, as the mass in stars grew, Bondi--Hoyle--Lyttleton accretion became the
more significant mechanism.
The transition occurred first for the most massive stars which had sunk
into the cluster core.
However, massive stars are very luminous and \citet{2004MNRAS.349..678E} showed
that radiative feedback can disrupt the Bondi--Hoyle--Lyttleton flow once
stellar masses exceed \unit{10}{\Msol} (this is obviously dependent on the
prevailing conditions in the protocluster).

Unfortunately, direct observations of this process are not available.
Protoclusters contain large quantities of dusty gas, which greatly obscure
regions of interest.
Furthermore, the expected luminosities are lower, and the emission wavelengths
less distinctive than those of X-ray binaries.

\subsection{Galaxy Clusters}

Galaxies orbiting in a cluster are another candidate for Bondi--Hoyle--Lyttleton
accretion.
One immediate complication is the high temperature of the intergalactic medium
(IGM).
The IGM is typically hot enough to emit X-rays, and hence the galactic motions
will usually be subsonic.
\citet{1971ApJ...165....1R} suggested that the IGM might be heated (at least in
part) by the accretion shocks inherent to Bondi--Hoyle--Lyttleton accretion.
Galaxies are also rather porous objects, and contain their own gas.
In a study of M86, \citet{1995MNRAS.277.1047R} concluded that the `plume' observed
was probably the result of ram-pressure stripping of material from the galaxy itself.
\citet{1999MNRAS.310..663S} simulated galaxies under such conditions, and
concluded that ``the ram-pressure stripped tail will usually be
the most visible feature.''
This paper also contains a list of observed wakes.

\citet{1980ApJ...242..511D} observed M87, and found evidence for \emph{subsonic}
Bondi--Hoyle--Lyttleton flow.
However, higher resolution observations by \citet{2000ApJ...543..611O} suggest
that this simple picture is not sufficient.
In particular, the active nucleus of M87 drives an outflow.
A filament has been observed trailing Abell 1795 both in the optical
\citep{1983ApJ...272...29C} and in X-rays
\citep{2001MNRAS.321L..33F}.
It has been proposed \citep{1996MNRAS.283..673S} that this filament is an
accretion wake, but \citet{2001MNRAS.321L..33F} note that the gas
cooling times aren't quite right for this simple approximation to be completely
valid.
\citet{2000MNRAS.318.1164S} constructed a simple theoretical model of a
Bondi--Hoyle--Lyttleton wake behind a galaxy.
Wakes were expected to extend for up to \unit{20}{\kilo\parsec}.
The wakes would form behind slow moving, massive galaxies in low temperature
clusters.

\subsection{Other Applications}

The original application of Bondi--Hoyle--Lyttleton
accretion was to the flow of the interstellar medium
past the Sun.
\citet{2002PhRvD..66b3504S} apply a similar analysis
to the flow of dark matter past the Sun.
They suggest that annual variations in WIMP detections
may be partially attributable to the focusing of
the flow by the Sun.

Bondi--Hoyle--Lyttleton accretion was invoked by
\citet{2002MNRAS.335L..45K} to explain the unusual chemical
abundances of $\lambda\mbox{-Bootis}$ type stars.
These stars have metal abundances typical of the interstellar
medium (that is, metal--poor).
\citeauthor{2002MNRAS.335L..45K} suggest that radiation pressure
on dust grains (which contain most of the metals) prevents the
accretion of the heavier elements, while gas accretes in
a Bondi--Hoyle--Lyttleton fashion.
Moving to higher energies, accretion onto neutron stars moving
through gas clouds has been proposed as a mechanism for producing
X-ray sources in the Galaxy \citep{1970ApL.....6..179O} and
in globular clusters \citep{2001ApJ...550..172P}.
However with neutron stars, magnetic fields can cause significant
complications - see, e.g. \citet{2001ApJ...561..964T}.

\citet{2002ApJ...570..671M} studied the central portions
of our Galaxy with Chandra.
Finding evidence for recent activity, they suggest that
this could have been powered by the central black hole
accreting material from an expanding supernova shock.
This would be a transient example of Bondi--Hoyle-Lyttleton
accretion.
The potential luminosity from this is rather high (comparable
with the Eddington Limit).
However, there is a complication due to the thermal pressure
of the ambient gas, which could reduce the inferred
luminosity substantially.

\section{Summary}

In their original analyses, Bondi, Hoyle and Lyttleton made many
simplifications.
Despite these, the broad picture they present seems to be correct.
Numerical studies have been made of the purely hydrodynamic problem, and
of cases where extra physical processes are relevant.
Bondi--Hoyle--Lyttleton accretion has also been used to explain
phenomena in a variety of astronomical contexts.

Of course, the original Bondi--Hoyle--Lyttleton results cannot be applied
without some thought.
Numerical studies have shown that the flow pattern is more complicated
than that originally envisaged.
Meanwhile, real systems are always more complicated than theoretical
ones.
Bondi--Hoyle--Lyttleton accretion should be regarded as a reference
model - it is unlikely to explain any system in detail, but it can
serve as a useful basis for classifying behaviours.
It can be applied as a test model on systems of all scales - from
binary stars up to galaxies in clusters.

There are many future avenues for research.
As well as improving simulations of accretion in binaries, studies
need to be made on Bondi--Hoyle--Lyttleton accretion for flows
where the accretor is orders of magnitude smaller than the
accretion radius.
Previous numerical work has hinted at lurking instabilities, but
sufficient computing resources have yet to be brought to bear.
Extra physical processes are also candidates for inclusion,
especially radiative feedback.
Radiative feedback is likely to be relevant in the context of
both X-ray binaries and star formation simulations, and could
alter the flow pattern significantly.
On the observational side, better observations of systems
previously modelled with a simple Bondi--Hoyle--Lyttleton analysis
will show deviations, which can then be used to enhance our
understanding of them.

Time has not eroded the usefulness of the Bondi--Hoyle--Lyttleton
accretion geometry.
It provides a simple framework for examining and refining theory
and observation.


\begin{ack}
I am indebted to my supervisor, Cathie Clarke, for her help and patience throughout my
PhD.
This work was completed while I was at Stockholm Observatory, funded by the EU--RTN Planets (HPRN-CT-2002-00308).
Matthew Bate and Jim Pringle both gave useful pointers to references for this article.
The derivation in section~\ref{sec:BondiHoyleAnalysis} is based on that
given by Douglas Gough in his lectures on Fluid Dynamics.
Gerry Gilmore read an early draft of this article, and made a number of helpful suggestions.
Finally, I am also grateful to the referees, for drawing further issues and work in the area to my attention.
\end{ack}


\bibliography{bibs/radiativetrans,bibs/dust,bibs/zeus,bibs/observations,bibs/hydro,bibs/compute,bibs/stellarevolve,bibs/accretiondisks,bibs/starform,bibs/bondihoyle}

\begin{thebibliography}{}

\bibitem[\protect\astroncite{{Binney} and
  {Tremaine}}{1987}]{1987gady.book.....B}
{Binney}, J. and {Tremaine}, S.: 1987,
\newblock {\em {Galactic dynamics}},
\newblock Princeton, NJ, Princeton University Press, 1987, 747 p.

\bibitem[\protect\astroncite{{Bisnovatyi-Kogan}
  et~al.}{1979}]{1979SvA....23..201B}
{Bisnovatyi-Kogan}, G.~S., {Kazhdan}, Y.~M., {Klypin}, A.~A., {Lutskii}, A.~E.,
  and {Shakura}, N.~I.: 1979,
\newblock {\em Soviet Astronomy} {\bf 23}, 201

\bibitem[\protect\astroncite{{Blondin} et~al.}{1990}]{1990ApJ...356..591B}
{Blondin}, J.~M., {Kallman}, T.~R., {Fryxell}, B.~A., and {Taam}, R.~E.: 1990,
\newblock {\em \apj} {\bf 356}, 591

\bibitem[\protect\astroncite{{Bondi}}{1952}]{1952MNRAS.112..195B}
{Bondi}, H.: 1952,
\newblock {\em \mnras} {\bf 112}, 195

\bibitem[\protect\astroncite{{Bondi} and {Hoyle}}{1944}]{1944MNRAS.104..273B}
{Bondi}, H. and {Hoyle}, F.: 1944,
\newblock {\em \mnras} {\bf 104}, 273

\bibitem[\protect\astroncite{{Bonnell} et~al.}{2001}]{2001MNRAS.323..785B}
{Bonnell}, I.~A., {Bate}, M.~R., {Clarke}, C.~J., and {Pringle}, J.~E.: 2001,
\newblock {\em \mnras} {\bf 323}, 785

\bibitem[\protect\astroncite{{Chandrasekhar}}{1943}]{1943ApJ....97..255C}
{Chandrasekhar}, S.: 1943,
\newblock {\em \apj} {\bf 97}, 255

\bibitem[\protect\astroncite{{Chapman}}{1981}]{1981ApJ...248.1043C}
{Chapman}, R.~D.: 1981,
\newblock {\em \apj} {\bf 248}, 1043

\bibitem[\protect\astroncite{{Che-Bohnenstengel} and
  {Reimers}}{1986}]{1986A&A...156..172C}
{Che-Bohnenstengel}, A. and {Reimers}, D.: 1986,
\newblock {\em \aap} {\bf 156}, 172

\bibitem[\protect\astroncite{{Cowie}}{1977}]{1977MNRAS.180..491C}
{Cowie}, L.~L.: 1977,
\newblock {\em \mnras} {\bf 180}, 491

\bibitem[\protect\astroncite{{Cowie} et~al.}{1983}]{1983ApJ...272...29C}
{Cowie}, L.~L., {Hu}, E.~M., {Jenkins}, E.~B., and {York}, D.~G.: 1983,
\newblock {\em \apj} {\bf 272}, 29

\bibitem[\protect\astroncite{{Davies} and
  {Pringle}}{1980}]{1980MNRAS.191..599D}
{Davies}, R.~E. and {Pringle}, J.~E.: 1980,
\newblock {\em \mnras} {\bf 191}, 599

\bibitem[\protect\astroncite{{De Young} et~al.}{1980}]{1980ApJ...242..511D}
{De Young}, D.~S., {Butcher}, H., and {Condon}, J.~J.: 1980,
\newblock {\em \apj} {\bf 242}, 511

\bibitem[\protect\astroncite{{Dodd} and {McCrea}}{1952}]{1952MNRAS.112..205D}
{Dodd}, K.~N. and {McCrea}, W.~J.: 1952,
\newblock {\em \mnras} {\bf 112}, 205

\bibitem[\protect\astroncite{{Dokuchaev}}{1964}]{1964SvA.....8...23D}
{Dokuchaev}, V.~P.: 1964,
\newblock {\em Soviet Astronomy} {\bf 8}, 23

\bibitem[\protect\astroncite{{Eadie} et~al.}{1975}]{1975MNRAS.172P..35E}
{Eadie}, G., {Peacock}, A., {Pounds}, K.~A., {Watson}, M., {Jackson}, J.~C.,
  and {Hunt}, R.: 1975,
\newblock {\em \mnras} {\bf 172}, 35P

\bibitem[\protect\astroncite{{Edgar} and {Clarke}}{2004}]{2004MNRAS.349..678E}
{Edgar}, R. and {Clarke}, C.: 2004,
\newblock {\em \mnras} {\bf 349}, 678

\bibitem[\protect\astroncite{{Fabian} et~al.}{2001}]{2001MNRAS.321L..33F}
{Fabian}, A.~C., {Sanders}, J.~S., {Ettori}, S., {Taylor}, G.~B., {Allen},
  S.~W., {Crawford}, C.~S., {Iwasawa}, K., and {Johnstone}, R.~M.: 2001,
\newblock {\em \mnras} {\bf 321}, L33

\bibitem[\protect\astroncite{{Fefferman}}{2000}]{NavierStokesUnique}
{Fefferman}, C.~L.: 2000,
\newblock in {\em Millenium Prize Problems}, p.~1, Clay Mathematics Institute

\bibitem[\protect\astroncite{{Foglizzo} and
  {Ruffert}}{1997}]{1997A&A...320..342F}
{Foglizzo}, T. and {Ruffert}, M.: 1997,
\newblock {\em \aap} {\bf 320}, 342

\bibitem[\protect\astroncite{{Foglizzo} and
  {Ruffert}}{1999}]{1999A&A...347..901F}
{Foglizzo}, T. and {Ruffert}, M.: 1999,
\newblock {\em \aap} {\bf 347}, 901

\bibitem[\protect\astroncite{{Foglizzo} and
  {Tagger}}{2000}]{2000A&A...363..174F}
{Foglizzo}, T. and {Tagger}, M.: 2000,
\newblock {\em \aap} {\bf 363}, 174

\bibitem[\protect\astroncite{{Font} and {Ib{\' a}{\~
  n}ez}}{1998a}]{1998MNRAS.298..835F}
{Font}, J.~A. and {Ib{\' a}{\~ n}ez}, J.~M.: 1998a,
\newblock {\em \mnras} {\bf 298}, 835

\bibitem[\protect\astroncite{{Font} et~al.}{1999}]{1999MNRAS.305..920F}
{Font}, J.~A., {Ib{\' a}{\~ n}ez}, J.~M., and {Papadopoulos}, P.: 1999,
\newblock {\em \mnras} {\bf 305}, 920

\bibitem[\protect\astroncite{{Font} and {Ib{\' a}{\~
  n}ez}}{1998b}]{1998ApJ...494..297F}
{Font}, J.~A. and {Ib{\' a}{\~ n}ez}, J.~M.~A.: 1998b,
\newblock {\em \apj} {\bf 494}, 297

\bibitem[\protect\astroncite{{Frank} et~al.}{2002}]{2002apa..book.....F}
{Frank}, J., {King}, A., and {Raine}, D.~J.: 2002,
\newblock {\em {Accretion Power in Astrophysics: Third Edition}},
\newblock Cambridge University Press

\bibitem[\protect\astroncite{{Fryxell} and {Taam}}{1988}]{1988ApJ...335..862F}
{Fryxell}, B.~A. and {Taam}, R.~E.: 1988,
\newblock {\em \apj} {\bf 335}, 862

\bibitem[\protect\astroncite{{Horedt}}{2000}]{2000ApJ...541..821H}
{Horedt}, G.~P.: 2000,
\newblock {\em \apj} {\bf 541}, 821

\bibitem[\protect\astroncite{{Hoyle} and {Lyttleton}}{1939}]{1939PCPS.34..405}
{Hoyle}, F. and {Lyttleton}, R.~A.: 1939,
\newblock {\em Proc.~Cam.~Phil.~Soc.} {\bf 35}, 405

\bibitem[\protect\astroncite{{Hunt}}{1971}]{1971MNRAS.154..141H}
{Hunt}, R.: 1971,
\newblock {\em \mnras} {\bf 154}, 141

\bibitem[\protect\astroncite{{Hunt}}{1979}]{1979MNRAS.188...83H}
{Hunt}, R.: 1979,
\newblock {\em \mnras} {\bf 188}, 83

\bibitem[\protect\astroncite{{Iben} and {Livio}}{1993}]{1993PASP..105.1373I}
{Iben}, I.~J. and {Livio}, M.: 1993,
\newblock {\em \pasp} {\bf 105}, 1373

\bibitem[\protect\astroncite{{Illarionov} and
  {Sunyaev}}{1975}]{1975A&A....39..185I}
{Illarionov}, A.~F. and {Sunyaev}, R.~A.: 1975,
\newblock {\em \aap} {\bf 39}, 185

\bibitem[\protect\astroncite{{Jackson}}{1975}]{1975MNRAS.172..483J}
{Jackson}, J.~C.: 1975,
\newblock {\em \mnras} {\bf 172}, 483

\bibitem[\protect\astroncite{{Kamp} and {Paunzen}}{2002}]{2002MNRAS.335L..45K}
{Kamp}, I. and {Paunzen}, E.: 2002,
\newblock {\em \mnras} {\bf 335}, L45

\bibitem[\protect\astroncite{{Kley} et~al.}{1995}]{1995A&A...297..739K}
{Kley}, W., {Shankar}, A., and {Burkert}, A.: 1995,
\newblock {\em \aap} {\bf 297}, 739

\bibitem[\protect\astroncite{{Koide} et~al.}{1991}]{1991MNRAS.252..473K}
{Koide}, H., {Matsuda}, T., and {Shima}, E.: 1991,
\newblock {\em \mnras} {\bf 252}, 473

\bibitem[\protect\astroncite{{Lamers} et~al.}{1976}]{1976A&A....49..327L}
{Lamers}, H.~J.~G.~L.~M., {van den Heuvel}, E.~P.~J., and {Petterson}, J.~A.:
  1976,
\newblock {\em \aap} {\bf 49}, 327

\bibitem[\protect\astroncite{{Livio} et~al.}{1991}]{1991MNRAS.253..633L}
{Livio}, M., {Soker}, N., {Matsuda}, T., and {Anzer}, U.: 1991,
\newblock {\em \mnras} {\bf 253}, 633

\bibitem[\protect\astroncite{{MacDonald}}{1980}]{1980MNRAS.191..933M}
{MacDonald}, J.: 1980,
\newblock {\em \mnras} {\bf 191}, 933

\bibitem[\protect\astroncite{{Maeda} et~al.}{2002}]{2002ApJ...570..671M}
{Maeda}, Y., {Baganoff}, F.~K., {Feigelson}, E.~D., {Morris}, M., {Bautz},
  M.~W., {Brandt}, W.~N., {Burrows}, D.~N., {Doty}, J.~P., {Garmire}, G.~P.,
  {Pravdo}, S.~H., {Ricker}, G.~R., and {Townsley}, L.~K.: 2002,
\newblock {\em \apj} {\bf 570}, 671

\bibitem[\protect\astroncite{{Matsuda} et~al.}{1987}]{1987MNRAS.226..785M}
{Matsuda}, T., {Inoue}, M., and {Sawada}, K.: 1987,
\newblock {\em \mnras} {\bf 226}, 785

\bibitem[\protect\astroncite{{Matsuda} et~al.}{1991}]{1991A&A...248..301M}
{Matsuda}, T., {Sekino}, N., {Sawada}, K., {Shima}, E., {Livio}, M., {Anzer},
  U., and {Boerner}, G.: 1991,
\newblock {\em \aap} {\bf 248}, 301

\bibitem[\protect\astroncite{{Nelson} and {Benz}}{2003}]{2003ApJ...589..578N}
{Nelson}, A.~F. and {Benz}, W.: 2003,
\newblock {\em \apj} {\bf 589}, 578

\bibitem[\protect\astroncite{{Ostriker} et~al.}{1970}]{1970ApL.....6..179O}
{Ostriker}, J.~P., {Rees}, M.~J., and {Silk}, J.: 1970,
\newblock {\em \aplett} {\bf 6}, 179

\bibitem[\protect\astroncite{{Owen} et~al.}{2000}]{2000ApJ...543..611O}
{Owen}, F.~N., {Eilek}, J.~A., and {Kassim}, N.~E.: 2000,
\newblock {\em \apj} {\bf 543}, 611

\bibitem[\protect\astroncite{{Petrich} et~al.}{1989}]{1989ApJ...336..313P}
{Petrich}, L.~I., {Shapiro}, S.~L., {Stark}, R.~F., and {Teukolsky}, S.~A.:
  1989,
\newblock {\em \apj} {\bf 336}, 313

\bibitem[\protect\astroncite{{Petterson}}{1978}]{1978ApJ...224..625P}
{Petterson}, J.~A.: 1978,
\newblock {\em \apj} {\bf 224}, 625

\bibitem[\protect\astroncite{{Pfahl} and
  {Rappaport}}{2001}]{2001ApJ...550..172P}
{Pfahl}, E. and {Rappaport}, S.: 2001,
\newblock {\em \apj} {\bf 550}, 172

\bibitem[\protect\astroncite{{Pfahl} et~al.}{2002}]{2002ApJ...571L..37P}
{Pfahl}, E., {Rappaport}, S., and {Podsiadlowski}, P.: 2002,
\newblock {\em \apjl} {\bf 571}, L37

\bibitem[\protect\astroncite{{Pounds} et~al.}{1975}]{1975MNRAS.172..473P}
{Pounds}, K.~A., {Cooke}, B.~A., {Ricketts}, M.~J., {Turner}, M.~J., and
  {Elvis}, M.: 1975,
\newblock {\em \mnras} {\bf 172}, 473

\bibitem[\protect\astroncite{{Rangarajan} et~al.}{1995}]{1995MNRAS.277.1047R}
{Rangarajan}, F.~V.~N., {White}, D.~A., {Ebeling}, H., and {Fabian}, A.~C.:
  1995,
\newblock {\em \mnras} {\bf 277}, 1047

\bibitem[\protect\astroncite{{Ruderman} and
  {Spiegel}}{1971}]{1971ApJ...165....1R}
{Ruderman}, M.~A. and {Spiegel}, E.~A.: 1971,
\newblock {\em \apj} {\bf 165}, 1

\bibitem[\protect\astroncite{{Ruffert}}{1994a}]{1994ApJ...427..342R}
{Ruffert}, M.: 1994a,
\newblock {\em \apj} {\bf 427}, 342

\bibitem[\protect\astroncite{{Ruffert}}{1994b}]{1994A&AS..106..505R}
{Ruffert}, M.: 1994b,
\newblock {\em \aaps} {\bf 106}, 505

\bibitem[\protect\astroncite{{Ruffert}}{1995}]{1995A&AS..113..133R}
{Ruffert}, M.: 1995,
\newblock {\em \aaps} {\bf 113}, 133

\bibitem[\protect\astroncite{{Ruffert}}{1996}]{1996A&A...311..817R}
{Ruffert}, M.: 1996,
\newblock {\em \aap} {\bf 311}, 817

\bibitem[\protect\astroncite{{Ruffert}}{1997}]{1997A&A...317..793R}
{Ruffert}, M.: 1997,
\newblock {\em \aap} {\bf 317}, 793

\bibitem[\protect\astroncite{{Ruffert}}{1999}]{1999A&A...346..861R}
{Ruffert}, M.: 1999,
\newblock {\em \aap} {\bf 346}, 861

\bibitem[\protect\astroncite{{Ruffert} and {Anzer}}{1995}]{1995A&A...295..108R}
{Ruffert}, M. and {Anzer}, U.: 1995,
\newblock {\em \aap} {\bf 295}, 108

\bibitem[\protect\astroncite{{Ruffert} and
  {Arnett}}{1994}]{1994ApJ...427..351R}
{Ruffert}, M. and {Arnett}, D.: 1994,
\newblock {\em \apj} {\bf 427}, 351

\bibitem[\protect\astroncite{{Sakelliou}}{2000}]{2000MNRAS.318.1164S}
{Sakelliou}, I.: 2000,
\newblock {\em \mnras} {\bf 318}, 1164

\bibitem[\protect\astroncite{{Sakelliou} et~al.}{1996}]{1996MNRAS.283..673S}
{Sakelliou}, I., {Merrifield}, M.~R., and {McHardy}, I.~M.: 1996,
\newblock {\em \mnras} {\bf 283}, 673

\bibitem[\protect\astroncite{{Shankar} et~al.}{1993}]{1993A&A...274..955S}
{Shankar}, A., {Kley}, W., and {Burkert}, A.: 1993,
\newblock {\em \aap} {\bf 274}, 955

\bibitem[\protect\astroncite{{Shapiro} and
  {Lightman}}{1976}]{1976ApJ...204..555S}
{Shapiro}, S.~L. and {Lightman}, A.~P.: 1976,
\newblock {\em \apj} {\bf 204}, 555

\bibitem[\protect\astroncite{{Shima} et~al.}{1998}]{1998A&A...337..311S}
{Shima}, E., {Matsuda}, T., {Anzer}, U., {Boerner}, G., and {Boffin}, H.~M.~J.:
  1998,
\newblock {\em \aap} {\bf 337}, 311

\bibitem[\protect\astroncite{{Shima} et~al.}{1985}]{1985MNRAS.217..367S}
{Shima}, E., {Matsuda}, T., {Takeda}, H., and {Sawada}, K.: 1985,
\newblock {\em \mnras} {\bf 217}, 367

\bibitem[\protect\astroncite{{Sikivie} and {Wick}}{2002}]{2002PhRvD..66b3504S}
{Sikivie}, P. and {Wick}, S.: 2002,
\newblock {\em \prd} {\bf 66}, 23504

\bibitem[\protect\astroncite{{Soker}}{1990}]{1990ApJ...358..545S}
{Soker}, N.: 1990,
\newblock {\em \apj} {\bf 358}, 545

\bibitem[\protect\astroncite{{Soker}}{1991}]{1991ApJ...376..750S}
{Soker}, N.: 1991,
\newblock {\em \apj} {\bf 376}, 750

\bibitem[\protect\astroncite{{Stevens} et~al.}{1999}]{1999MNRAS.310..663S}
{Stevens}, I.~R., {Acreman}, D.~M., and {Ponman}, T.~J.: 1999,
\newblock {\em \mnras} {\bf 310}, 663

\bibitem[\protect\astroncite{{Struck} et~al.}{2004}]{2004MNRAS.347..173S}
{Struck}, C., {Cohanim}, B.~E., and {Anne Willson}, L.: 2004,
\newblock {\em \mnras} {\bf 347}, 173

\bibitem[\protect\astroncite{{Taam} and
  {Bodenheimer}}{1989}]{1989ApJ...337..849T}
{Taam}, R.~E. and {Bodenheimer}, P.: 1989,
\newblock {\em \apj} {\bf 337}, 849

\bibitem[\protect\astroncite{{Taam} and {Fryxell}}{1988}]{1988ApJ...327L..73T}
{Taam}, R.~E. and {Fryxell}, B.~A.: 1988,
\newblock {\em \apjl} {\bf 327}, L73

\bibitem[\protect\astroncite{{Taam} et~al.}{1991}]{1991ApJ...371..696T}
{Taam}, R.~E., {Fu}, A., and {Fryxell}, B.~A.: 1991,
\newblock {\em \apj} {\bf 371}, 696

\bibitem[\protect\astroncite{{Taam} and
  {Sandquist}}{2000}]{2000ARA&A..38..113T}
{Taam}, R.~E. and {Sandquist}, E.~L.: 2000,
\newblock {\em \araa} {\bf 38}, 113

\bibitem[\protect\astroncite{{Theuns} et~al.}{1996}]{1996MNRAS.280.1264T}
{Theuns}, T., {Boffin}, H.~M.~J., and {Jorissen}, A.: 1996,
\newblock {\em \mnras} {\bf 280}, 1264

\bibitem[\protect\astroncite{{Theuns} and
  {Jorissen}}{1993}]{1993MNRAS.265..946T}
{Theuns}, T. and {Jorissen}, A.: 1993,
\newblock {\em \mnras} {\bf 265}, 946

\bibitem[\protect\astroncite{{Toropina} et~al.}{2001}]{2001ApJ...561..964T}
{Toropina}, O.~D., {Romanova}, M.~M., {Toropin}, Y.~M., and {Lovelace},
  R.~V.~E.: 2001,
\newblock {\em \apj} {\bf 561}, 964

\bibitem[\protect\astroncite{{Wang}}{1981}]{1981A&A...102...36W}
{Wang}, Y.-M.: 1981,
\newblock {\em \aap} {\bf 102}, 36

\bibitem[\protect\astroncite{{Wolfson}}{1977a}]{1977ApJ...213..200W}
{Wolfson}, R.: 1977a,
\newblock {\em \apj} {\bf 213}, 200

\bibitem[\protect\astroncite{{Wolfson}}{1977b}]{1977ApJ...213..208W}
{Wolfson}, R.: 1977b,
\newblock {\em \apj} {\bf 213}, 208

\bibitem[\protect\astroncite{{Yabushita}}{1978}]{1978MNRAS.182..371Y}
{Yabushita}, S.: 1978,
\newblock {\em \mnras} {\bf 182}, 371

\end{thebibliography}
\bibliographystyle{astron}

\end{document}